\documentclass{aa}  
\usepackage{pdflscape}  
\usepackage{longtable}
\usepackage{booktabs}   
\usepackage{pdfpages}
\usepackage{arydshln}
\usepackage{multirow}
\usepackage{graphicx}
\usepackage{txfonts}
\usepackage{hyperref}
\usepackage{natbib}
\bibpunct{(}{)}{;}{a}{}{,} 
\usepackage{natbib}
\begin{document}
\title{A FLASH on blazars}
\subtitle{Capturing the radio realm of 4FGL blazars with SKA Observatory pathfinders}
\titlerunning{A FLASH on blazars}

\author{{M. Behiri }\inst{1}\fnmsep\inst{2}\fnmsep\inst{3}\thanks{\email{mbehiri@sissa.it}} \and E. K. Mahony\inst{4} \and E. Sadler\inst{4}\fnmsep\inst{5}\fnmsep\inst{6} \and E. Kerrison\inst{4}\fnmsep\inst{5}\fnmsep\inst{6}  \and A. Traina\inst{7} \and M. V. Zanchettin\inst{8} \and V. Galluzzi\inst{2} \and A. Lapi\inst{1}\fnmsep\inst{2}\fnmsep\inst{3}\fnmsep\inst{9} \and M. Massardi\inst{10}\fnmsep\inst{1}}
\institute{Scuola Internazionale Superiore di Studi Avanzati, Via Bonomea 265, 34136 Trieste, Italy \and INAF - Istituto di Radioastronomia, Via Gobetti 101, 40129 Bologna, Italy \and IFPU - Institute for fundamental physics of the Universe, Via Beirut 2, 34014 Trieste, Italy \and  ATNF, CSIRO Space and Astronomy, PO Box 76, Epping, NSW 1710, Australia \and Sydney Institute for Astronomy, School of Physics A28, University of Sydney, NSW 2006, Australia \and ARC Centre of Excellence for All Sky Astrophysics in 3 Dimensions (ASTRO 3D) \and INAF-OAS Bologna, via Gobetti 101, I-40129 Bologna, Italy \and INAF - Osservatorio Astrofisico di Arcetri, Largo E. Fermi 5, I-50125, Firenze, Italy\and INFN, Sezione di Trieste, via Valerio 2, Trieste I-34127, Italy \and INAF, Istituto di Radioastronomia, Italian ARC, Via Piero Gobetti 101, I-40129 Bologna, Italy}



\abstract 
{
This work investigates the multi-wavelength properties of 165 sources from the \textit{Fermi} Large Area Telescope Fourth Source Catalog Data Release 4 (4FGL), with the aim of identifying counterparts in the Australian SKA Pathfinder (ASKAP) First Large Absorption Survey in the HI (FLASH) continuum. Using high-resolution data from FLASH, complementary radio datasets, and archival ALMA observations, we performed detailed spectral energy distribution (SED) analyses across centimetre to millimetre wavelengths. Our findings reveal that most blazars exhibit re-triggered peaked spectra, which are indicative of emission dominated by a single emitting region. Additionally, we identify strong correlations between radio and gamma-ray luminosities, highlighting the significant role of relativistic jets in these active galactic nuclei. The inclusion of spectroscopic redshifts from the Sloan Digital Sky Survey and Gaia enabled a comprehensive analysis of the evolutionary trends and physical characteristics of the sources to be performed. Furthermore, we report a tight radio--X-ray correlation for flat spectrum radio quasars, thus contrasting with the more scattered behaviour observed in BL Lacertae objects and reflecting the distinct accretion and jet-driving mechanisms of the two populations. These results provide critical insights into the physics of blazars and their environments, paving the way for future studies with next-generation facilities such as the SKA Observatory for radio observations and the Cherenkov Telescope Array for gamma-ray studies.
}
\maketitle



\section{Introduction} \label{sec:intro}
The latest radio interferometers are opening up a brand new perspective on the main characters of the radio sky: active galactic nuclei (AGN). Observatories such as the Murchison Widefield Array (MWA; \citealt{tingay13}) and the Australian SKA Pathfinder (ASKAP) are able to target faint populations at low radio frequencies, efficiently covering wide areas of the sky. The large multi-wavelength statistics of the objects surveyed by these telescopes finally makes it possible to answer several open questions on the physics and the evolutionary role of the different AGNs types.
\begin{figure*}
    \centering
   \includegraphics[width=17cm]{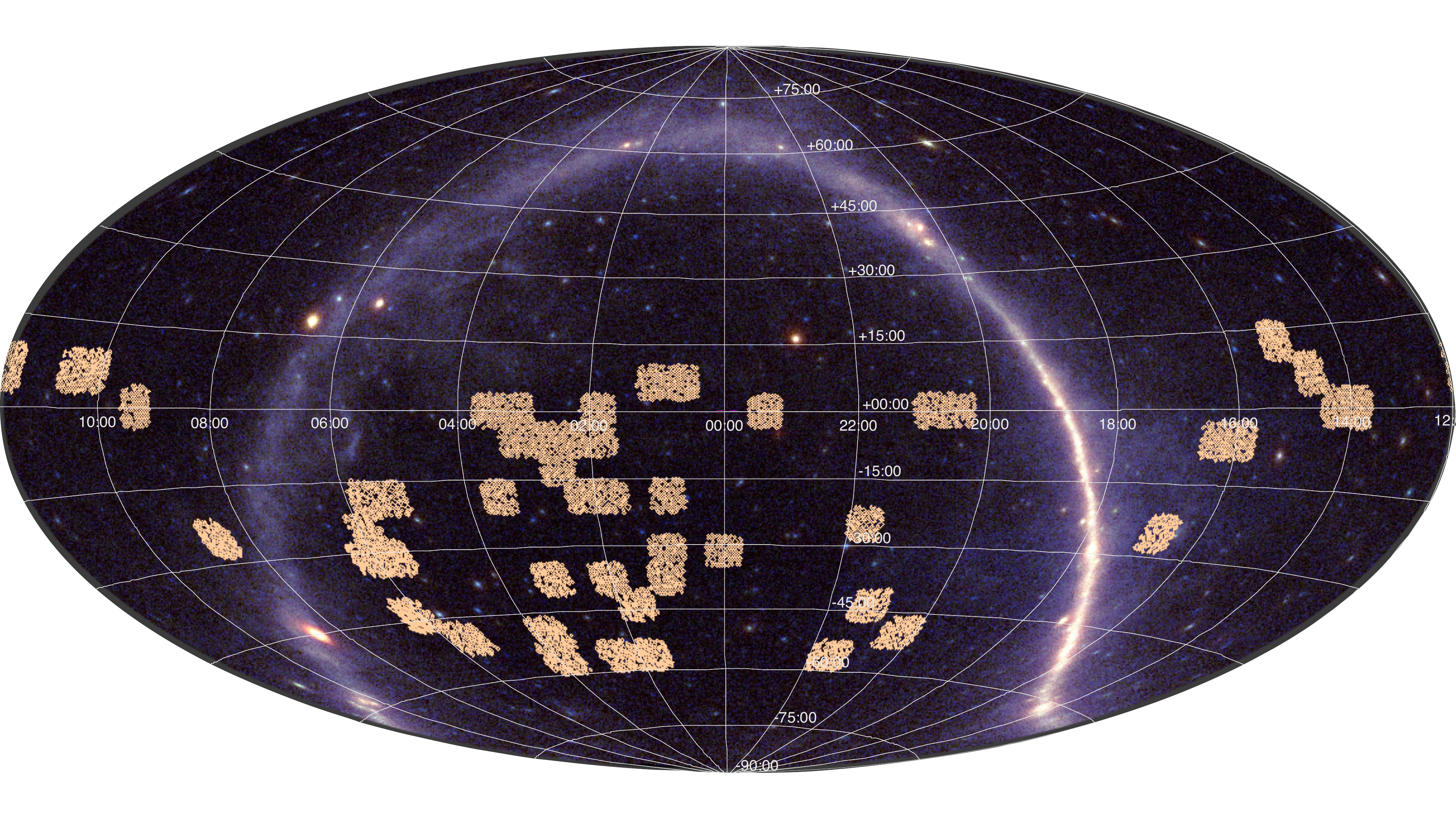}
    \caption{Map showing the coverage of \textit{Fermi}-LAT 4FGL. The light pink regions represent the Multi-Order Coverage of the FLASH fields utilised in this work.}
    \label{fig:moc}
\end{figure*}

Within the radio spectrum, the primary emission process is synchrotron radiation, which results from electrons moving at relativistic speeds as they spiral around magnetic field lines. This process is characteristic of radio AGNs, and it produces their distinctive relativistic jets. Through the analysis of the radio spectral energy distribution (SED) of AGNs, it is possible to retrieve clues about their history. For instance, an inverted spectrum is an indication of self-absorption processes, which are typical of compact sources. On the other end, a break at high radio frequency suggests the presence of ageing processes, while a flat spectrum could either be an indication of efficient energy pumping continuously refurnishing the jets with energetic electrons or of the superposition along the line of sight of several components. Further, changes in the trend of the SED might be a proxy for reactivation episodes of the AGN. Thus, broad radio spectral coverage is key to reaching a comprehensive view of the radio emission of AGNs.

Among the several classes of AGNs, blazars appear rather interesting from a radio point of view. According to the Unified Model by \cite{urrypadovani}, blazars are AGNs whose jets are oriented almost ($<$ 15-20 deg) along the line of sight. They are characterised by Doppler boosting of their flux density, apparent superluminal motion, large and rapid variability, and strong non-thermal emission over the entire electromagnetic spectrum \citep{padovani16}. \begin{figure*}[]
    \centering
    \includegraphics[width=0.3\linewidth]{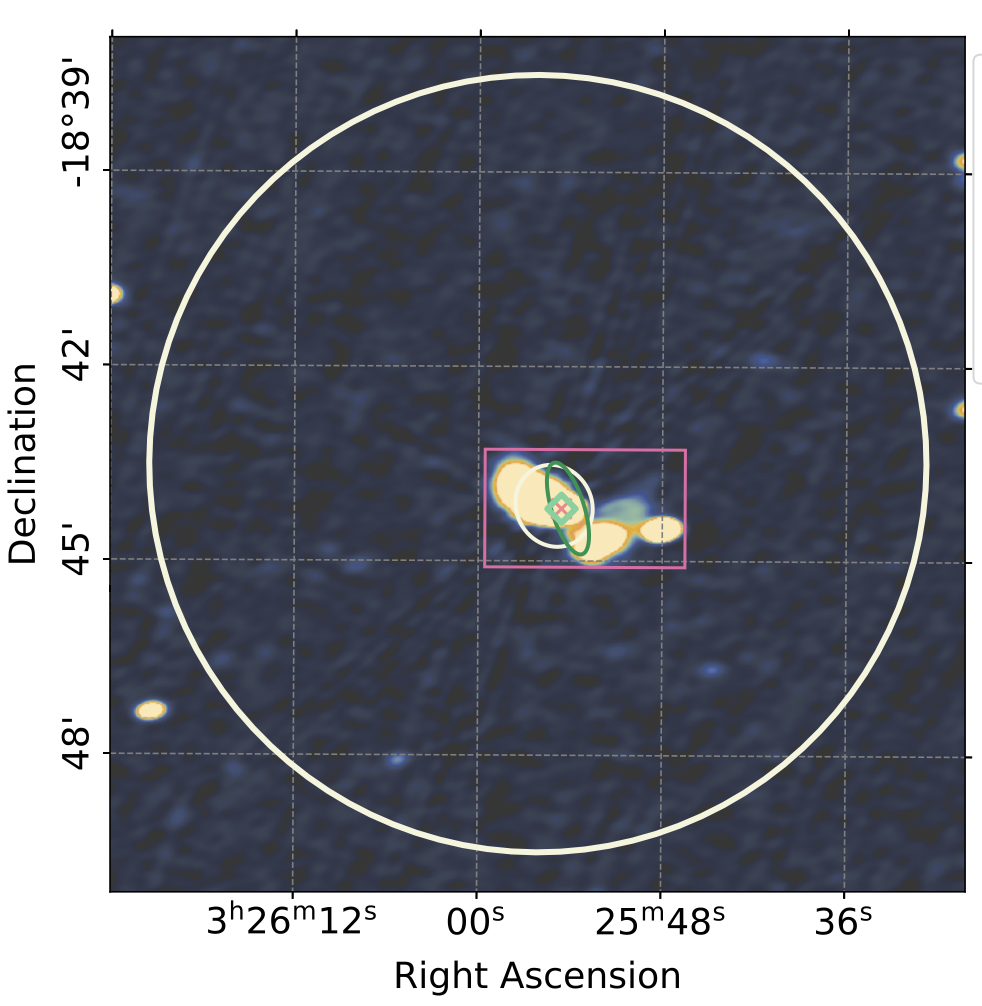}\quad\includegraphics[width=0.3\linewidth]{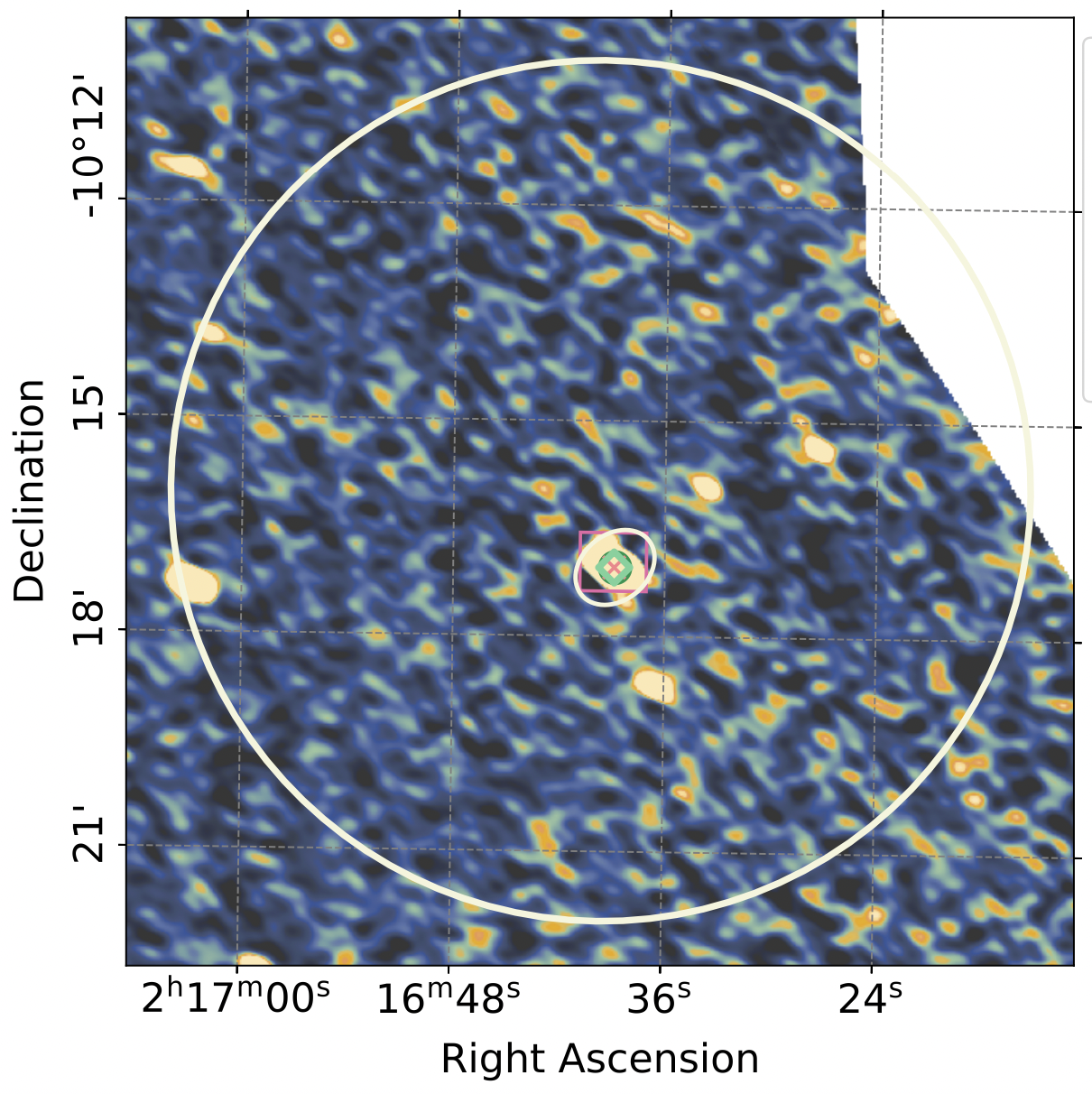}\quad\includegraphics[width=0.3\linewidth]{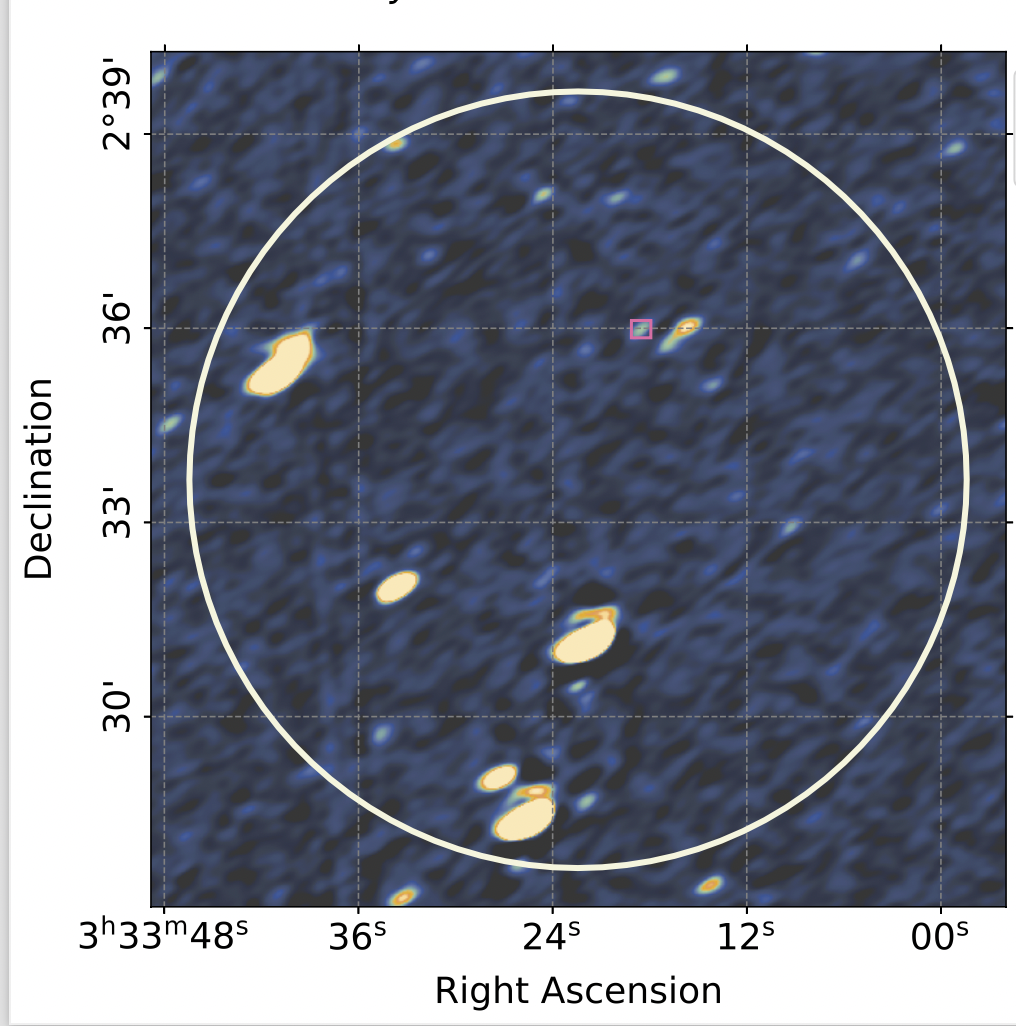}
    \caption{Illustrations showing three sources categorised as `IN' \textit{(Left)},`OUT' \textit{(Center)}, and `UNCERTAIN' \textit{(Right)}, respectively. These images are based on FLASH maps. The large white circle is aligned with the \textit{Fermi}-LAT location, sharing the same radius as the \textit{Fermi}-LAT's resolution (0.1 degrees). The pink rectangle is positioned at the FLASH continuum island's location, with dimensions matching those of the island as determined by \textsc{Selavy} \citep{selavy}. The dark green ellipse indicates the size and position of the corresponding RACS-LOW island, and the small white circle marks the GLEAM-X matched source's location, with a size equivalent to the (median) GLEAM-X resolution (in arcseconds). The pink `X' and light green diamond denote the best-corresponding positions of the AT20G source to the 4FGL position and the FLASH continuum position, respectively. }
    \label{fig:selection}
\end{figure*}

Blazars constitute an ideal background target for detecting HI absorption in both the interstellar medium (ISM) and intergalactic medium, as their intense radio emission enables the detection of absorption at high redshifts, which can help improve our understanding of the distribution and properties of neutral gas in the Universe (e.g. \citealt{klemola,HIabs}). Thus, projects such as the First Large Absorption Survey in HI (FLASH; \citealt{flash}), made with ASKAP, leverage the characteristics of blazars to study HI absorption, neutral gas evolution, and jet-environment interactions. The ability of blazars to illuminate cosmologically distant environments makes them indispensable tools for investigating the distribution of neutral gas and its interaction with relativistic jets. Therefore, ahead of the upcoming data from FLASH and the completion of this survey, a first step towards the full accomplishment of using blazars for HI absorption would be to understand the peculiarities of the blazar population that FLASH is able to target. Understanding the radio properties of blazars in the FLASH continuum catalogue is key to exploiting their potential as astrophysical probes. In particular, their strong gamma-ray emission makes gamma-ray selection an effective way to identify and classify them \citep{abdo10,nolan12,acero15,giroletti+16}. The \textit{Fermi} Large Area Telescope Fourth Source Catalog Data Release 4 (4FGL) by \cite{4fgl} offers the perfect playground for identifying blazars. Blazars are strong gamma-ray emitters through inverse Compton scattering of the jet electrons with external photons, typically from the disc. They can also produce gamma-rays via synchrotron self-Compton emission from the same electrons responsible for the radio emission, and through hadronic processes. Hadronic processes are generated by protons losing energy through synchrotron emission (if the magnetic field is large enough) or photo-meson reactions. In the latter case, neutrino emission from the decaying charged pions is also expected \citep{tavecchio2017,padovani22a,padovani22b}.

In the 4FGL catalogue, blazars are divided into two subclasses. One of the classes is {BL Lacertae objects (BL Lac)}, which are characterised by extremely weak or even absent emission lines in their optical spectra. In these objects, gamma-ray emission is predominantly driven by synchrotron self-Comptonisation. Their AGN emission is thought to be mainly driven by the rotation of the supermassive black hole. The second class is {flat spectrum radio quasars (FSRQs)}, which exhibit prominent emission lines in their optical spectra. In the gamma-ray regime, their emission is predominantly due to inverse Comptonisation. These sources are considered to be accretion driven and extremely powerful across the spectrum, displaying the typical flat spectrum in the radio. 

The link between these two classes has been highly debated within the framework of the Unified Model. It has been suggested that an FSRQ could evolve into a BL Lac as it loses its power, leaving a rotationally driven structure \citep{cavaliere2002}.
Traditionally, all blazars were thought to be characterised by a flat radio spectrum. However, some studies addressing the issue with multi-wavelength data (e.g. \citep{massardi+16}) have already highlighted that the situation may have varying nuances. \cite{massardi+16} showed that about 90\% of the blazars in their sample have a remarkably smooth spectrum well described by a power law over the range 5–150 GHz. This suggests that the emission in this frequency interval is dominated by a single emitting region rather than by the superposition of different compact regions self-absorbed up to different frequencies. \cite{giroletti+16} drew a scenario that differs from the traditional view by combining MWA Commissioning Survey (\citealt{mwacs}) observations at 180 MHz and gamma-ray emission. They showed that extended steep spectrum emission might still be present in quasars at low frequencies, thus confirming the possible presence of a jet.

Nevertheless, to further explore the correlation between radio and gamma-ray emission, deep high-resolution radio panchromatic coverage is needed. The advent of wide-area low-frequency surveys, such as the GaLactic and Extragalactic All-sky Murchison Widefield Array eXtended (GLEAM-X, \citealt{mwa}) and the ASKAP surveys, have extended the radio coverage of the southern sky to lower frequencies. In particular, the FLASH continuum survey is significant in the endeavour to conduct detailed studies of synchrotron emission and its connection to gamma-ray sources, as the data from this survey offer an unexplored perspective on the low-energy emission of blazars. 

In this work, we investigate the radio emission of 4FGL blazars detected in the FLASH continuum survey \citep{flash}, reconstructing their SEDs using multi-frequency radio data from FLASH, the Rapid ASKAP Continuum Survey (RACS; \citealt{racs1,racs2}), GLEAM-X \citep{mwa}, the Australian Telescope compact array 20 GHz (AT20G; \citealt{at20g1}) survey, and the Atacama Large Millimeter Array CALibrators catalogue (ALMACAL;\citealt{almacal}). 
We analyse their spectral trends, explore the presence of re-triggered spectral features, and study correlations with gamma-ray and X-ray emissions. 
Our findings provide new insights into blazar jet properties, their evolutionary states, and their role as cosmological tracers of neutral gas distribution. These results will play a fundamental role in projects such as FLASH that leverage the characteristics of blazars to study HI absorption, neutral gas evolution, and jet-environment interactions. 
This study also lays the groundwork for future investigations with the SKA Observatory and the Cherenkov Telescope Array, extending blazar studies into the next-generation radio and gamma-ray era.
\section{Sample selection} \label{sec:samp_sel}
\begin{table}[h]
    \caption{Surveys used in this study.}
    \label{tab:catalogues}
    \centering
    \begin{tabular}{|c|c|c|}
    \hline
Survey & \# Matches  & Spectral coverage \\
\hline
4FGL &- & 100 MeV-1 TeV \\
eROSITA & 65& 0.2-10 keV \\
QUAIA & 77 & 380-1080 nm  \\
SDSS &30 &u,g,r,i,z  \\
ALMACAL  &21 & 95-350 GHz\\
AT20G  & 32 & 4, 8, 20 GHz  \\
RACS &165 & 887.5-2367.5 MHz\\
FLASH & 165 &711.5–999.5 MHz   \\
GLEAM-X  &97 &72-231 MHz\\
\hline
    \end{tabular}
\end{table}
The starting point for our sample selection was the 4FGL catalogue, DR4 \citep{4fgl}. Our sample consists of the BL Lac, FSRQs, and blazar candidates of uncertain type (BCUs) from this catalogue with counterparts in the FLASH continuum maps. In particular, we used the 50 FLASH maps flagged as `good', which were downloaded up to November 21, 2024 (Figure \ref{fig:moc}). Since the survey is still ongoing, the number of publicly available `good; fields may have increased since then.
In this section, we outline the primary catalogues used and detail the criteria for selecting the sources for our sample. The surveys used in this work are listed in Table \ref{tab:catalogues}.
\subsection{FLASH}
{FLASH is a radio survey conducted with ASKAP \citep{flash}}. It is designed to detect neutral hydrogen in absorption against bright radio continuum sources, and probes the diffuse and dense phases of the intergalactic and ISM over cosmological scales. The survey covers a redshift range of $0.4 < z < 1.0$. 
In this work, we make use of the FLASH continuum observations. FLASH continuum observations cover the frequency range 711.5–999.5 MHz, with a typical sensitivity of $\sim$0.2 mJy/beam and an angular resolution of $\sim$15''. 
\subsection{4FGL}
The 4FGL represents the most comprehensive survey in the 100 MeV–1 TeV energy range. It is based on eight years of \textit{Fermi}/LAT mission data and includes over 5000 sources. Each source is characterised by its position; gamma-ray flux (including photon flux, energy flux, flux in different energy bands, and temporal variability across 96 monthly time bins); and spectral properties such as the photon index. The typical 95$\%$ positional confidence radius is approximately $\sim$ 0.05–0.1 deg, depending on the source's brightness and location in the sky \citep{4fgl}.
\subsection{Selection process}
We performed a cross-match between the two catalogues using a radius of 0.1 deg, which corresponds to the full width at half maximum of LAT, as between the two instruments, LAT possesses the least effective angular separation since the ASKAP beam size is of $\sim$ 15 arcsec at 855 MHz (FLASH central frequency).
Given the large difference in the resolving power of the surveys, we visually inspected the data to exclude spurious and ambiguous pairings (Figure \ref{fig:selection}).

As FLASH is composed of isolated mosaics (Figure \ref{fig:moc}), some of the sources happened to be on the edge of the fields, where the reliability drops down. In some cases, more than one FLASHsource was equally probable to be the counterpart of one FERMI source. In the second case, we did not directly exclude the sources. Instead, we took advantage of other existing radio surveys, namely AT20G \citep{at20g1,at20g2}, RACS \citep{racs1}, and GLEAM-X \citep{mwa}, to decide, via visual inspection, which FLASH source was the most suitable counterpart. Nevertheless, for some objects it was impossible to determine which was the more likely counterpart, and thus we classified them as `UNCERTAIN'.
We obtained a total of 259 cross-matched sources, and of these, we classified 278 as IN, 33 as OUT, and 80 as UNCERTAIN.

\subsection{The spectroscopic sub-sample}
\begin{figure}[h]
    \centering
    \resizebox{\hsize}{!}{\includegraphics{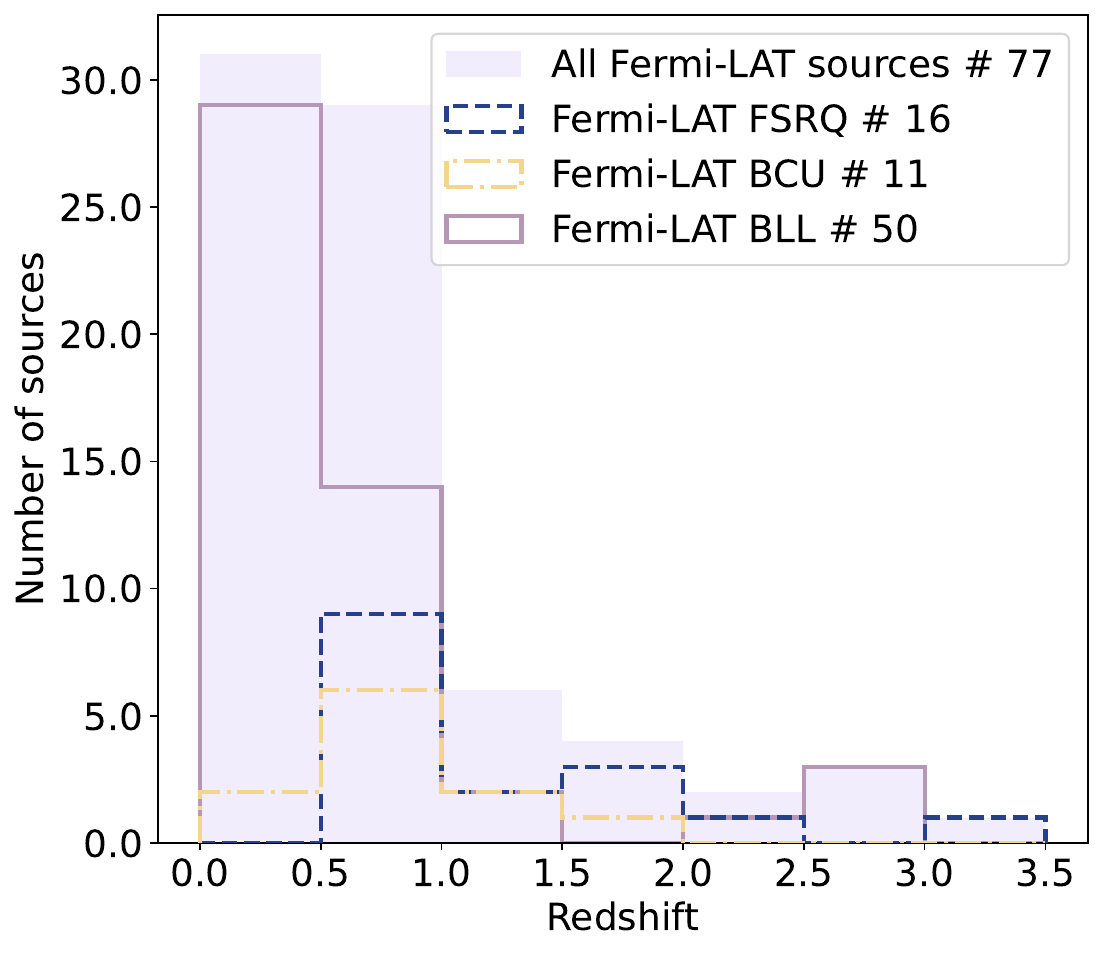}}
    \caption{Distribution of spectroscopic redshifts from SDSS for sources marked as 'IN' with a spectroscopic counterpart.}
    \label{fig:redshift_dist}
\end{figure}

To find the spectroscopic redshifts of the sources, we cross-matched the FLASH positions of the sources in our sample with the Sloan Digital Sky Survey (SDSS) DR17 survey and the Gaia–unWISE Quasar Catalog (QUAIA), using a radius of 5 arcsec.
SDSS covers $\sim$ 14000 deg$^2$ of the sky, with a typical photometric sensitivity of 22.5 magnitudes in five optical bands (u, g, r, i, z) and a spatial resolution of $\sim$ 1.4 arcsec, providing spectra of the objects with a resolution of $R \sim $ 2000 at 3600–10400 $\AA$. SDSS overlaps FLASH in the equatorial area.
QUAIA is an all-sky spectroscopic quasar sample. The catalogue draws on the 6,649,162 quasar candidates identified by the Gaia mission that have redshift estimates from the space observatory’s low-resolution blue photometer/red photometer spectra.  As a result, we achieved a total of 77 spectroscopically confirmed objects. The median redshift of these sources is 0.77 and thus consistent with the overall FLASH catalogue (\ref{fig:redshift_dist}). 
Figure \ref{fig:redshift_dist} illustrates that BL Lacs are generally at lower redshifts than FSRQs, bolstering the hypothesis of a parent-child relationship between the two populations. However, a larger spectroscopic sample is required to assess this scenario.
\section{Methods}
\subsection{Radio properties} \label{subsec:radio_prop}
\begin{table}
      \caption{Summary of the median radio properties of the 59 fitted sources from RadioSED.}
         \label{tab:radio_props}
  \centering
    \begin{tabular}{|c|c|}
    \hline
       Quantity &  Median Value \\
       \hline
       $\alpha_{\mathrm{retrig}}$ & -0.5\\ 
      $\alpha_{\mathrm{thick}}$ & 0.2\\
      $\alpha_{\mathrm{thin}}$ & -0.6\\
        $\nu_{\mathrm{peak-rest}}$ & 4.6 GHz\\
        $\nu_{\mathrm{trough-rest}}$ & 1.0 GHz \\
        \hline
    \end{tabular}
\end{table}

To fully reconstruct the radio SED of the sources and thus retrieve their properties, we used all the radio data available for the objects in our sample. A key aspect to consider when analysing multi-frequency data is the significant disparity in angular resolution across different surveys. At lower frequencies (e.g. GLEAM-X at $\sim$72-231 MHz), the resolution is $\sim$45'', while at 20 GHz (AT20G), it improves to a few arcseconds. This resolution mismatch could introduce uncertainties when associating extended radio structures with gamma-ray sources. However, since most blazars are expected to be compact at GHz frequencies, this effect is likely minimal.

We cross-matched the FLASH coordinates from our catalogue with GLEAM-X, RACS, AT20G, and ALMACAL \citep{almacal}, enabling us to achieve comprehensive photometric coverage across radio frequencies ranging from 72 MHz (GLEAM-X) to approximately 30 GHz (ALMACAL).
In this section, we outline the development of the radio multi-wavelength catalogue, detail the SED-fitting process, and present the results derived from this procedure.

The survey RACS covers the whole sky at $-90<\delta<49$ at $\sim$887.5/943.5 (RACS-low,\citealt{racs1,racs2}); 1367.5 (RACS-mid, \citealt{racsmid}); and 1655.5 MHz (RACS-high,\citealt{racshigh}), reaching a sensitivity of $\sim$0.25 (RACS-low) and 0.2 mJy/beam (RACS-mid, high) with a resolution of $\sim$ 15'' (RACS-low), 10'' (RACS-mid), and 8'' (RACS-high). Thus, RACS-low observations allowed us to explore the sub-gigahertz emission as well. We cross-matched the catalogue using a 15'' radius, and as expected, we found that all of our sources have a counterpart in RACS.

The GLEAM-X survey covers an area of 1447 deg$^2$ around the South Galactic Pole region covering $-32.7<\delta<-20.7$ with a median sensitivity of 6.35 mJy/beam and a median resolution of 45'' \citep{mwa}. The main advantage of GLEAM-X is that it is composed of twenty frequency bands in the range 72–231 MHz. This interval samples the megahertz side of the radio SED, where self-absorption processes occur.

{Regarding AT20G, it is a blind 20 GHz survey of the entire southern sky conducted with the Australian Telescope Compact Array (ATCA). It reaches a sensitivity of 40 mJy/beam \citep{at20g1} and a resolution of $\sim$ 0.15''. Further, most sources, $\delta <-15.0$, also have near-simultaneous measurements at 5 and 8 GHz. The primary advantage of AT20G for blazar SED studies is its high-frequency coverage and sensitivity over a large sky area \citep{at20g2}.}

{
The ALMACAL survey is constructed from reprocessed ALMA calibration scans suitable for scientific analysis \citep{almacal}. Thus, the fields are randomly distributed across the ALMA-visible sky, cumulatively covering $\sim$ 0.3 deg$^2$. Its frequency coverage spans from 95 to 350 GHz, making this catalogue extremely useful in the SED modelling of bright AGNs. }

\subsection*{Variability and temporal aspects of the data}
\label{sec:vardata}
{Blazars are intrinsically variable sources across the entire electromagnetic spectrum, including the radio band. Since our spectral analysis is based on multi-frequency archival data, we acknowledge that the non-simultaneity of observations may introduce uncertainties in the derived spectral shapes. However, most of our results—and the sample selection itself—are based on the very low frequency bands, where radio variability is estimated to remain below 5\% \citep{massardi+11}. At higher frequencies, variability increases significantly, but the lack of recent and deep wide-area millimetric surveys makes it impossible to obtain fully coeval observations. To account for this, we systematically increased the flux uncertainties for the AT20G and ALMACAL measurements. In the following, we provide a brief overview of the time sampling characteristics of the datasets we used to assess the potential impact of variability.}

{GLEAM-X \citep{mwacs} observed the 72–231\,MHz range with the Murchison Widefield Array between 2018 and 2020. The band was covered in five subbands, cycled through every $\sim$10 minutes, resulting in quasi-simultaneous coverage of the full range for each source.}
{RACS \citep{racs1} was conducted in separate campaigns for different frequencies: RACS-low (887.5 MHz) from April 2019 to mid 2020 and RACS-mid (1367.5 MHz) in a later epoch. Although the observations are not simultaneous, many sources have data in both bands within a temporal gap of $\lesssim$1.5 years.}

{FLASH \citep{flash} provides continuum fluxes in the 711.5–999.5 MHz range, observed simultaneously within each ASKAP pointing. For sources where both FLASH and RACS data are available, we used them cautiously, being aware of possible offsets due to variability.}

{AT20G \citep{at20g1,at20g2} was carried out between 2004 and 2008. Although the initial blind scan was done at 20 GHz, follow-up observations at 5 and 8 GHz were performed within a short time (weeks) of detection, offering a relatively consistent three-frequency snapshot.}

{ALMACAL \citep{almacal} spans from ALMA Cycle 0 (2011) to the present. While individual calibrator scans are short and of a single frequency, repeated observations across multiple cycles provide multi-frequency coverage. In this work, we used only fluxes taken within the same or adjacent ALMA cycles to limit the impact of long-term variability.}

{In summary, although our dataset is heterogeneous and was not selected to minimise variability, in many cases the time intervals between observations are sufficiently short to reduce the most significant inconsistencies. We discuss the possible implications of variability in the interpretation of our results in Section \ref{sec:var}.}

\subsection{Radio SED fitting}
\begin{figure}
    \centering
   \includegraphics[width=0.8\linewidth]{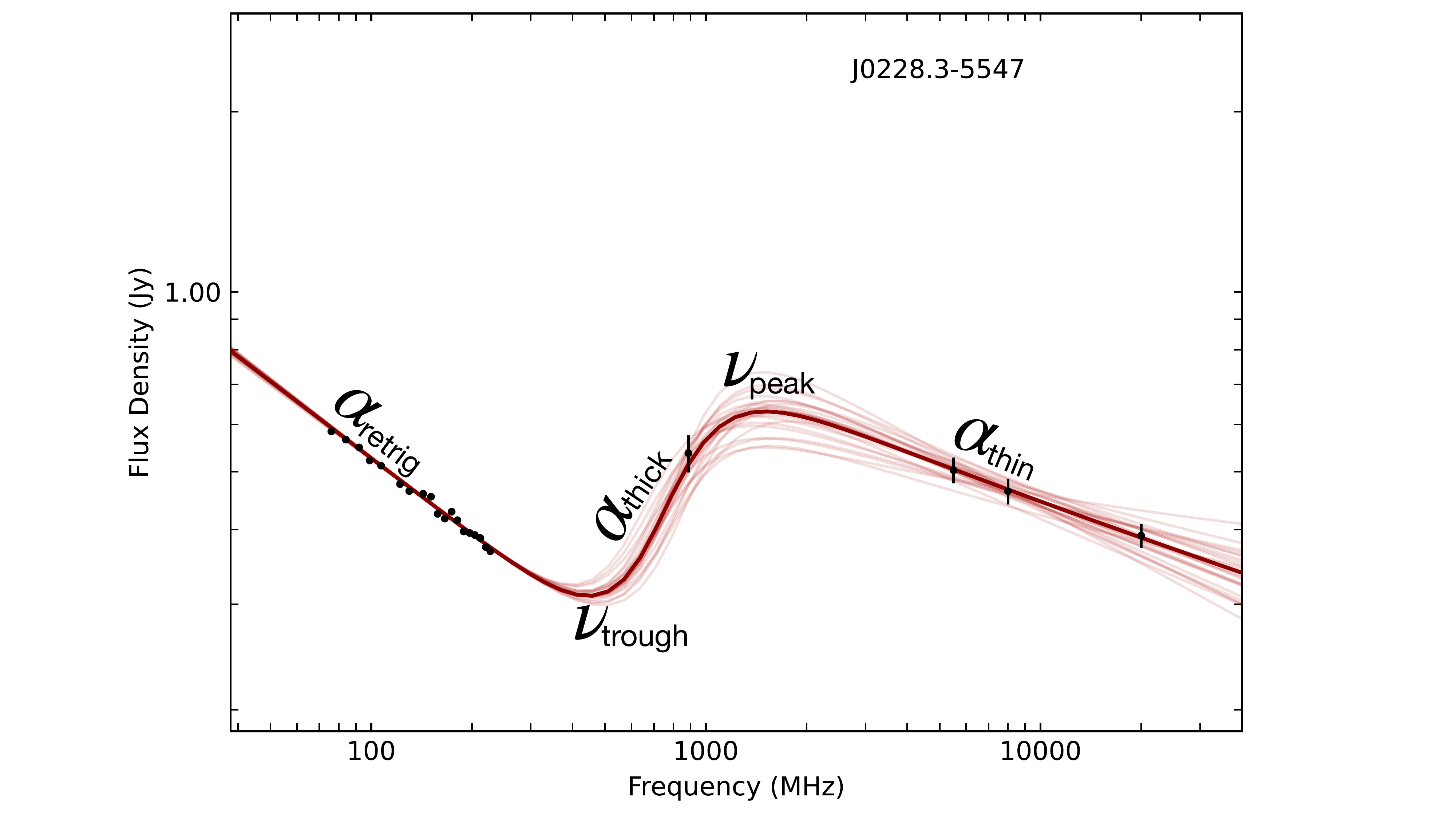}
 \caption{Spectral energy distribution fitting for one of the selected blazars in our sample. The models incorporate GLEAM-X, RACS, AT20G, and ALMACAL data. 
{The dark red line is the best fit from the posterior of the selected model type, while individual photometric points are marked. The light red lines are random draws from the posterior representing the range of best-fitting parameters.}}
 \label{fig:seds}
 \end{figure}

The SED fitting procedure included all the sources with at least five photometric detections in the radio bands.
We performed the SED fitting using \textsc{RADIOSED} \citep{radiosed}, software designed to model and fit radio SEDs of astrophysical sources (Figure \ref{fig:seds}, Appendix \ref{sec:appendix}).
This tool is particularly suited to analysing multi-frequency radio data, thus allowing us to disentangle the contributions of various emission mechanisms by testing multiple spectral models. 
\textsc{RADIOSED} supports different functional forms to describe radio SEDs. For this paper and consistent with \citet{radiosed}, we classified the spectral shapes as follows:
\begin{itemize}
    \item \textbf{Power law:} This class corresponds to a simple model of the form 
    \begin{equation}
        S_\nu \propto \nu^\alpha,
    \end{equation}
    where $\alpha$ is the spectral index. Flat spectrum sources, traditionally associated with blazar jets \citep{massaro13a,massaro13b}, fall into this category. 
    
    \item \textbf{Peaked spectrum:} This class includes sources that exhibit a turnover in their SED, where the flux density increases up to a peak frequency, $\nu_p$, and then decreases. This behaviour can indicate synchrotron self-absorption or a young radio source. We modelled these sources using the functional form \citep{snellen98,callingham,slob}
    \begin{equation}
        S_\nu = \frac{S_p}{1 - e^{-1}} \left( 1 - e^{-(\nu/\nu_p)^{\alpha_{\mathrm{thin}} - \alpha_{\mathrm{thick}}}} \right) \left( \frac{\nu}{\nu_p} \right)^{\alpha_{\mathrm{thick}}},
    \end{equation}
    where $S_p$ is the flux density at $\nu_p$, and $\alpha_{\mathrm{thin}}$ and $\alpha_{\mathrm{thick}}$ are the spectral indices for the optically thin and thick components.
    
    \item \textbf{Curved peaked spectrum:} This class represents a modification of the peaked spectrum that includes an additional curvature term to better capture broad or asymmetric spectral shapes \citep{orienti+07,orienti+10}:
    \begin{equation}
        \log S_{\nu}= a + \log \nu \left( b + c \log \nu \right),
    \end{equation}
    where $c$ is the curvature parameter.
    
    \item \textbf{Re-triggered peaked spectrum:} This class corresponds to a peaked SED with a secondary low-frequency upturn, suggesting episodic jet activity or restarted accretion \citep{edwards+04}. It is modelled as
    \begin{equation}
        S_\nu = \frac{S_p}{1-e^{-1}}\left[1 - e^{-(\nu/\nu_p)^{\alpha_{\mathrm{thin}}-\alpha_{\mathrm{thick}}}} \right]\left( \frac{\nu}{\nu_p} \right)^{\alpha_{\mathrm{thick}}} + S_0\nu^\alpha.
        \label{eq:retr}
    \end{equation}
\end{itemize}

Each model was fitted independently, and the best-fit model was determined through Bayesian model selection. \textsc{RadioSED} computes the Bayes factor to compare the likelihoods of different models, selecting the one with the highest posterior probability. 

{We adopted a standard Gaussian likelihood non-informative priors (log-uniform, linear-uniform), as described in Table 2 of \citet{radiosed}, to ensure a statistically robust classification of spectral types that provide a reliable characterisation of the blazar SEDs in our sample. }
However, the SED modelling does not account for the intrinsic variability between epochs, which can affect the interpretation of blazar spectra. This represents a limitation of our modelling approach that should be considered when analysing the results.
\section{Results}
\label{sec:sed}
From the SED fitting of the radio emission, we retrieved the spectral behaviour of the sources and found that most of them are well described by the re-triggered peaked spectrum as defined by Equation \ref{eq:retr}. This smooth behaviour, in contrast to a flat spectrum, suggests that the emission is dominated by a single emitting region rather than by the superposition of multiple compact, self-absorbed regions, as indicated by the classic blazar paradigm \citep{massardi+16}.

The main parameters retrieved from the SED fitting (Table \ref{tab:radio_props}, Figure \ref{fig:seds}) are the following:
\begin{itemize}
    \item $\nu_{\mathrm{trough}}$: This frequency corresponds to the minimum of the SED. After this point, if present, the SED starts to rise.
    \item $\alpha_{\mathrm{trough}}$: This is the spectral index at frequencies $\nu<\nu_{\mathrm{trough}}$, which usually corresponds to GLEAM-X photometric points for our sample.
    \item $\nu_{\mathrm{peak}}$: This is the frequency where, if present, the peak of the SED takes place. 
    \item  $\alpha_{\mathrm{thick}}$: This is the spectral index at frequencies $\nu_{\mathrm{trough}}<\nu<\nu_{\mathrm{peak}}$, which corresponds to the thick part of the peaked component.
    \item $\alpha_{\mathrm{thin}}$: This is the spectral index at frequencies $\nu>\nu_{\mathrm{peak}}$, which corresponds to the thin part of the peaked component.
\end{itemize}
Using the advanced surveys employed in this study, we conclude that at GLEAM-X wavelengths, the spectral behaviour is not completely flat ($\alpha \sim -0.5$, Table \ref{tab:radio_props}). This finding opens the path to an additional extended steep-spectrum emission that was previously missed \citep{massaro13a,massaro13b,nori14}. GLEAM-X is in fact sensitive to diffuse low-frequency emission likely from older radio lobes that are usually missed in higher-frequency surveys.
\begin{figure}[h]
    \centering
    \resizebox{\hsize}{!}{\includegraphics{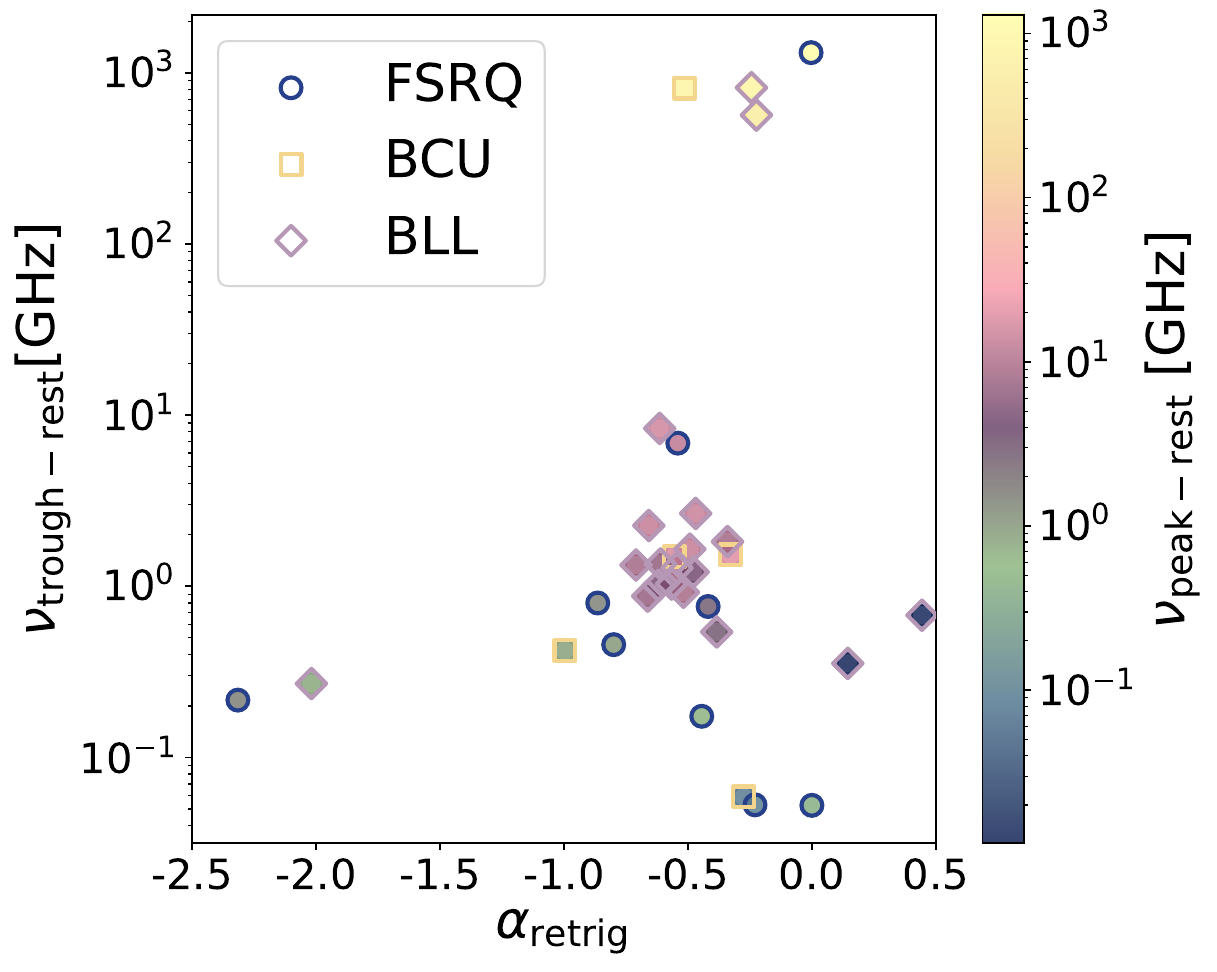}}
\caption{Relation between the trough frequency, $\nu_{\text{trough}}$, and the spectral index at lower frequencies, $\alpha_{\text{trough}}$. 
The colour scale represents the peak frequency, $\nu_{\text{peak}}$, which can be used as a proxy for the cooling time of the source. The colours of the markers' edges represent the three blazar populations: FSRQs, BL Lacs, and BCUs.}
    \label{fig:trough}
\end{figure}
\begin{figure}[h]
    \centering
    \includegraphics[width=0.8\linewidth]{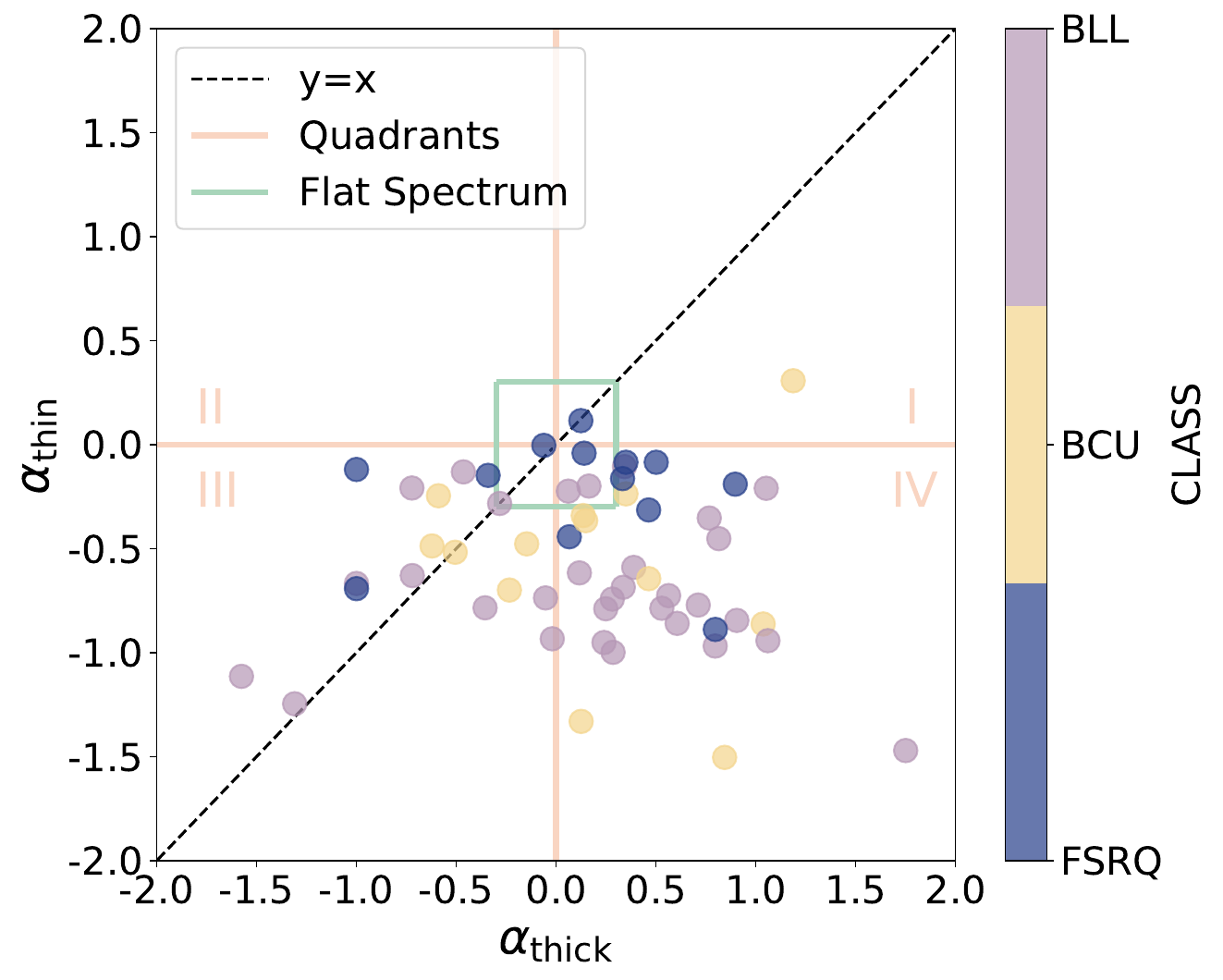}
\caption{Radio colour-colour diagram of the blazars in our sample based on the spectral indices retrieved from the SED fitting. 
The colours represent the three blazar populations: FSRQs, BL Lacs, and BCUs.}
    \label{fig:colour-colour}
\end{figure}

According to the theory that $t_{\mathrm{cool}} \propto 1/\sqrt{\nu_{\mathrm{peak}}}$, where $\nu_{\mathrm{peak}}$ serves as a proxy for evolutionary timescales, our sample reveals that younger objects tend to exhibit their spectral upturn at higher frequencies while maintaining a low-frequency spectral index of $\sim -0.5$ (Figure \ref{fig:trough}). Conversely, older sources are less constrained to a specific spectral index, but they all exhibit a spectral shape change around $\sim$100 MHz. This behaviour might indicate that a more powerful core might still be prominent and efficient in younger sources, efficiently refurnishing the jet with energetic electrons.

For most of the sources, the SED extends smoothly towards higher frequencies with a peaked spectrum emission, where it is possible to distinguish a thick part ($\alpha_{\mathrm{thick}}$) and a thin part ($\alpha_{\mathrm{thin}}$). 
The median $\nu_{\mathrm{peak rest}}$ is $\sim$ 4.6 GHz and no hint of breaks due to ageing is present at high frequency. 

In the radio colour-colour diagram, displayed in Figure \ref{fig:colour-colour}, it emerges that most of the objects ($\sim$ 57.6 \%) are in the fourth quadrant; that is, they have a hint of a peaked spectrum, and $\sim$ 10.2\% have a flat spectrum after $\nu_{\mathrm{trough}}$. This is a rather small percentage, and it indicates that in the gigahertz domain as well, we should expect a single emitting region rather than the superposition of a number of compact regions of different sizes self-absorbed at different frequencies.

\begin{figure*}[h!]
    \centering
    \includegraphics[width=17cm]{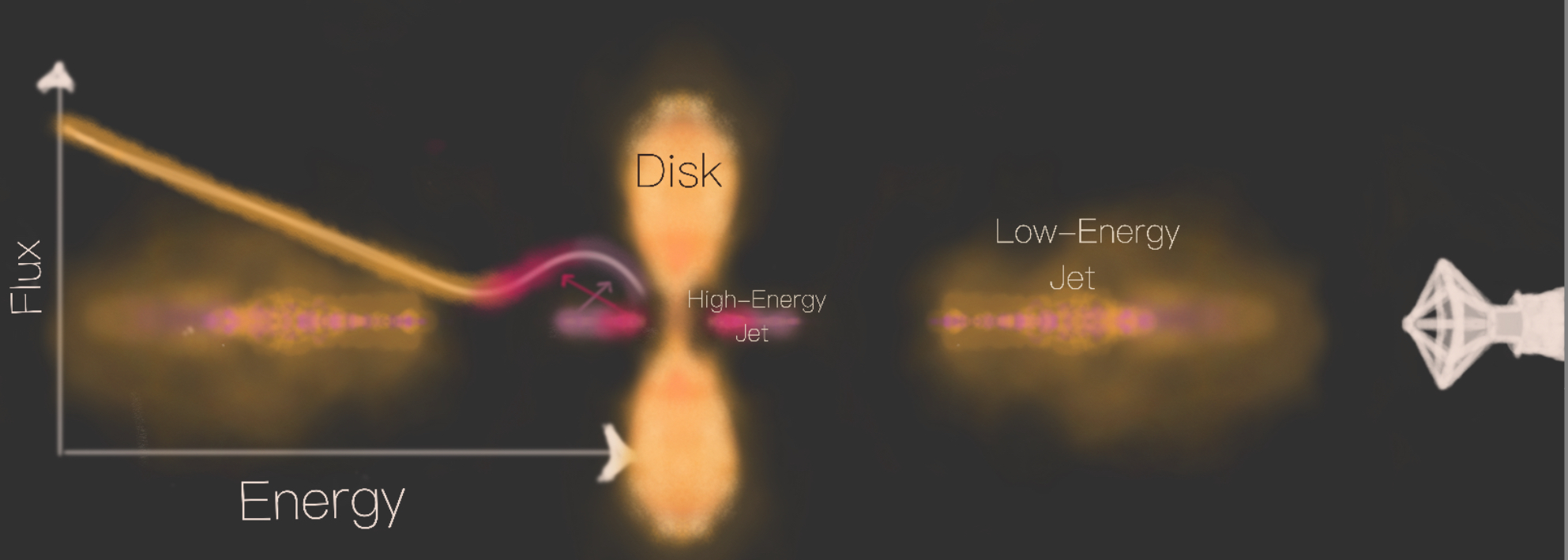}
   \caption{Schematic representation of the physical interpretation of blazar radio emission outlined in this paper. The figure illustrates the interplay between compact self-absorbed components and extended steep-spectrum emission, which may indicate episodic jet activity or restarted AGN cycles.}

    \label{fig:scheme}
\end{figure*}

This single component dominating at $\nu>\nu_{\mathrm{trough}}$ could represent the compact self-absorbed segment that was re-activated at a later time. Meanwhile, the low-energy segment of the SED could delineate an extended remnant from earlier activity (Figure \ref{fig:scheme}). Our results suggest that resolution differences have a limited impact on the main trends observed in the radio SEDs, as most sources are unresolved at gigahertz frequencies. Nevertheless, deeper high-resolution, low-frequency surveys (e.g. LOFAR, SKA-Low) will be valuable for disentangling diffuse extended emission from compact jet components.

On the radio colour-colour diagram (Figure \ref{fig:colour-colour}), BL Lacs display a wide range of behaviours, whereas FSRQs tend to cluster in the flat region. This indicates that FSRQs undergo highly efficient electron pumping, which significantly boosts high-energy radio synchrotron emissions. 

Conversely, the two populations seem indistinguishable when considering $\nu_{\mathrm{trough}}$ and $\alpha_{\mathrm{trough}}$ (Figure \ref{fig:colour-colour}).
This may suggest that the extremely low radio frequency emissions are remnants of the source's history, remaining unchanged by subsequent activities detected at higher frequencies, as the former emissions are confined to the outskirts, while the latter is linked to the nuclear core region.
\subsection*{On the impact of radio variability}
\label{sec:var}
{Due to the non-simultaneity of the multi-frequency radio data used in this work, some level of spectral distortion may result from intrinsic variability, especially in the most compact and core-dominated blazars. While we did not filter our sample based on observation dates, in many cases the frequency bands were observed within time spans short enough to reduce major inconsistencies (see Section \ref{sec:vardata}). Still, care must be taken in the interpretation of features such as spectral turnovers or curvature, particularly in the higher frequency domain, where variability tends to be more pronounced. In future work, dedicated multi-epoch or strictly simultaneous observations will help constrain the temporal evolution of blazar spectra further and disentangle variability-driven effects from intrinsic spectral shapes.}

\section{Discussion}
\subsection{The radio-gamma correlation}\label{sec:radio-gamma}
\begin{figure*}[ht!]
    \centering
  \includegraphics[width=17cm]{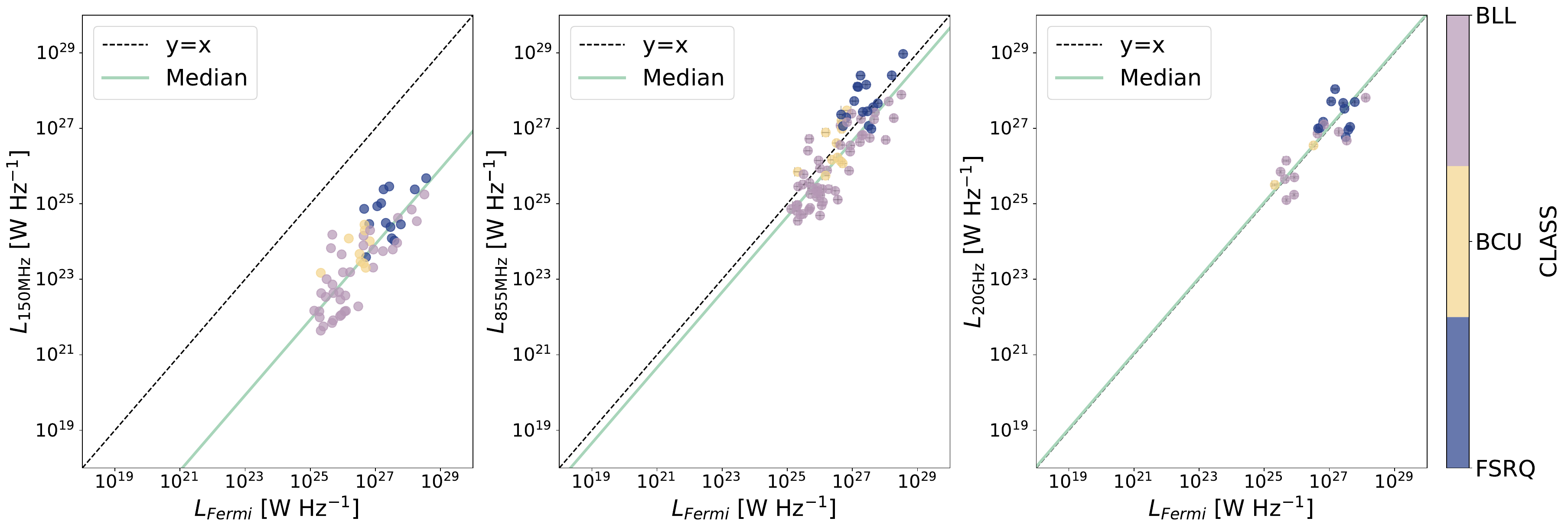}
        \caption{Radio luminosity at 150 MHz (\textit{left}), 855 MHz (\textit{centre}), and 20 GHz (\textit{right}) as a function of the $\gamma$-ray luminosity from \textit{Fermi}-LAT. The black dashed line and the green line indicate the bisector and the median trend, respectively. The colours indicate the blazar class from the 4FGL catalogue.}
    \label{fig:gamma-150}
\end{figure*}
\begin{table}
   \caption{Median and correlation between the ratio of radio and $\gamma$-ray luminosities for different object classes.}
    \label{tab:gamma_radio}
    \centering
    \begin{tabular}{|c|c|c|c|c|}
        \hline
          \multicolumn{2}{|c|}{}     & All & FSRQ & BL Lac \\ \hline
        \multirow{2}{*}{$\mathrm{log} \Big (\frac{L_{\mathrm{150MHz}}}{L_{\gamma}} \Big )$}  
        & Median & -3.07   & -2.82  & -3.25 \\ \cdashline{2-5}
        & Correlation & 0.70   & 0.77  & 0.93\\ \hline
        \multirow{2}{*}{$\mathrm{log} \Big (\frac{L_{\mathrm{855MHz}}}{L_{\gamma}} \Big )$}  
        & Median & 0.34   & 0.38  & -0.48 \\ \cdashline{2-5}
        & Correlation & 0.70   & 0.92  & 0.85\\ \hline
        \multirow{2}{*}{$\mathrm{log} \Big (\frac{L_{\mathrm{20GHz}}}{L_{\gamma}} \Big )$}  
        & Median & 0.06   & 0.15  & -0.14 \\ \cdashline{2-5}
        & Correlation & 0.78   & 0.20   & 0.98 \\ \hline
    \end{tabular}

\end{table}
Using the redshift estimates, we have drawn a picture of the energy budget spent by the sources in the radio bands from $\sim$ 72 MHz to $\sim$ 30 GHz and in the \textit{Fermi}-LAT bands. This combined with the radio SED fitting described in Section \ref{sec:sed} allowed us to find the primary emitting components of the object and its evolutionary stage.
The median value of the radio luminosity at 855 MHz is 2.5$\times10^{26}\mathrm{WHz^{-1}}$ for the spectroscopic sample, with 25\% of the sample having a luminosity greater than $1.8\times10^{27}\mathrm{WHz^{-1}}$.
In addition, 25\% of the sources have a $\gamma$-ray luminosity higher than $1.8\times10^{27}\mathrm{WHz^{-1}}$, and the median value is $4.3\times10^{26}\mathrm{WHz^{-1}}$.

The correlation between radio and $\gamma$-ray emission, if present, supports the idea that they are both dominated by the relativistic jet \citep{giroletti+16,mahony16}.
In this work, how this relation behaves in different radio bands can be observed (Figure \ref{fig:gamma-150}, Table \ref{tab:gamma_radio}). This is particularly useful since the dominant emitting component varies along the radio spectrum, as we found through the SED fitting.
The luminosity at 855 MHz is closely tied to the $\gamma$-ray luminosity, indicating that at this frequency, the jet significantly impacts both radio and gamma-ray energy emissions, with a tight relation (correlation coefficient $\rho\sim 0.70$, see Figure \ref{fig:gamma-150}). This suggests that at 855 MHz, we primarily observe the most powerful segment of the jet nearest the nucleus (Figure \ref{fig:scheme}), which potentially represents the re-activated section. 
This is supported by the way the radio--gamma-ray relation appears at 20GHz. Consequently, at high radio frequencies, the radio emission is overwhelmingly driven by a strong jet.

In Figure \ref{fig:gamma-150} the correlation between 150 MHz and $\gamma$ emission is displayed. Although in this case as well there is a correlation between high- and low-energy emissions (correlation coefficient $\sim0.70$), there is an evident offset. A potential explanation for this might be that the extended synchrotron emission, which is not efficient enough at generating $\gamma$-rays, dominates this part of the spectrum. This scenario is in full agreement with the results obtained from the SED fitting (Section \ref{sec:sed}).

The FSRQs dominate the bright end of the luminosity in both radio and $\gamma$, and the radio-gamma correlation holds for FSRQs and BL Lacs. Nevertheless, as shown in Table \ref{tab:gamma_radio}, the radio-gamma correlation assumes higher values for FSRQs at 150 and 855 MHz, which approaches the point at which the radio and $\gamma$ emission are perfectly balanced at lower frequencies. Thus, the FSRQs could already saturate the relation at lower frequencies. This suggests that FSRQs might generally exhibit more effective high-energy pumping across various scales and reflects the presence of a flat spectrum at $\nu>\nu_{trough}$; that is, the emitting mechanism at 855 and 20 GHz is almost the same. 
\subsection{The radio--X-ray correlation}
\begin{figure*}
    \centering
\includegraphics[width=17cm]{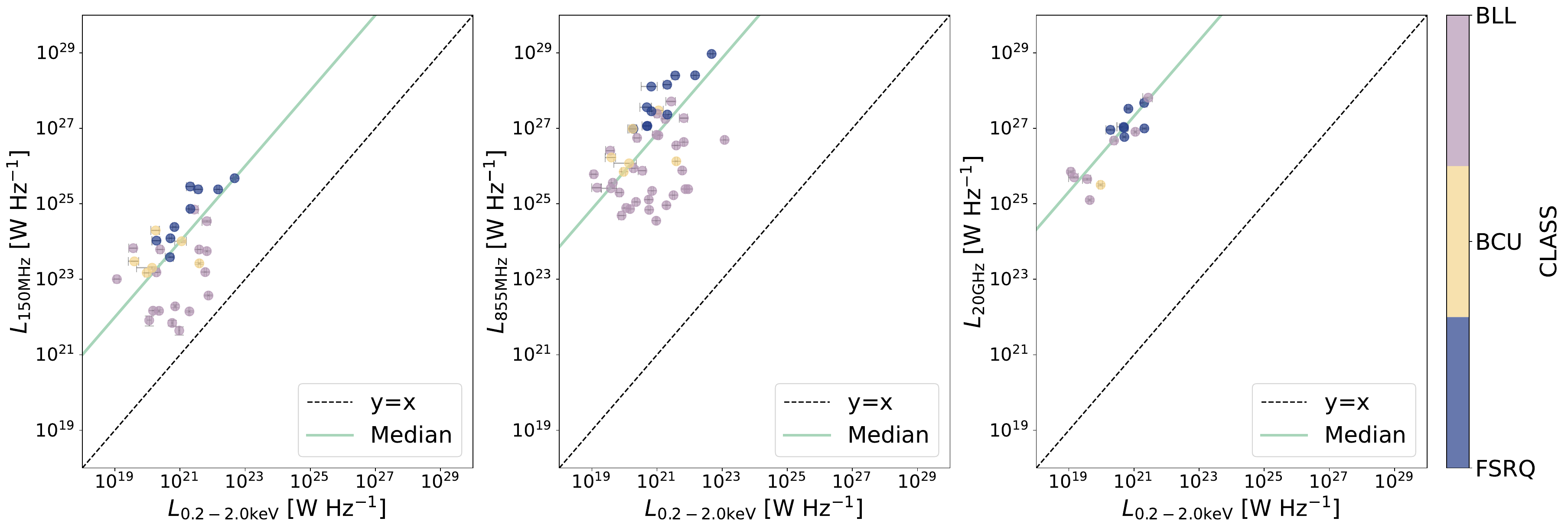}
    \caption{Radio luminosity at 150 MHz (\textit{left}), 855 MHz (\textit{centre}), and 20 GHz (\textit{right}) as a function of the X-ray luminosity from eROSITA. The black dashed line indicates the bisector. The colours indicate the blazar class from the 4FGL catalogue.}
    \label{fig:xray-radio}
\end{figure*}

To further explore the nature of blazars, we cross-matched our sample with the eROSITA all-sky survey \citep{eROSITA}. Forty-eight of the sources in the spectroscopic sample have an eROSITA counterpart within 10''. 
The eROSITA telescope array aboard the Spektrum Roentgen Gamma (SRG) satellite began surveying the sky at 0.2-10 keV. The eROSITA DR1 covers $\sim$ 14056 deg$^2$, with a median sensitivity in the 0.2-2 keV of $\sim$ 10$^{-14}$erg cm$^{-2}$s$^{-1}$. 
For the cross-matched sources, we were able to study the relationship between X-ray and radio emission.
We considered the energy range 0.2 - 2.0 keV for the X-ray luminosity. 
Within this spectrum, the analysed sources exhibit a median luminosity of 6.9$\times10^{20}\mathrm{WHz^{-1}}$, with 25\% of them having a luminosity exceeding $2.8\times10^{21}\mathrm{WHz^{-1}}$.
\begin{table}
    \caption{Median and correlation between the ratio of radio and X-ray luminosities for different object classes.}
    \label{tab:xray_radio}
    \centering
    \begin{tabular}{|c|c|c|c|c|}
        \hline
          \multicolumn{2}{|c|}{}     & All & FSRQ & BL Lac \\ \hline
        \multirow{2}{*}{$\mathrm{log} \Big (\frac{L_{\mathrm{150MHz}}}{L_{0.2-2.0\mathrm{keV}}} \Big )$}  
        & Median & 3.00   & 3.55   & 2.09 \\ \cdashline{2-5}
        & Correlation & 0.80    & 0.81  & 0.28 \\ \hline
        \multirow{2}{*}{$\mathrm{log} \Big (\frac{L_{\mathrm{855MHz}}}{L_{0.2-2.0\mathrm{keV}}} \Big )$}  
        & Median & 5.86    & 6.62  & 6.62 \\ \cdashline{2-5}
        & Correlation & 0.32   & 0.97  & 0.02 \\ \hline
        \multirow{2}{*}{$\mathrm{log} \Big (\frac{L_{\mathrm{20GHz}}}{L_{0.2-2.0\mathrm{keV}}} \Big )$}  
        & Median & 6.32   & 6.34   & 6.33 \\ \cdashline{2-5}
        & Correlation & 0.82  & 0.48   & 0.99 \\ \hline
    \end{tabular}
\end{table}
The main driver of the emission in this range is the hot corona, which closely traces the accretion disc, especially in FSRQs. The radio--X-ray correlation is shown in Figure \ref{fig:xray-radio}. We show the results in Table \ref{tab:xray_radio}, and we found almost no correlation for the 855 MHz but a tight one at 150 MHz.
In this particular situation, when even giving just a glance at Figure \ref{fig:xray-radio}, it is clear that the global behaviour might be misleading and not truly representative of the physical processes taking place. The real picture emerges only when distinguishing between different classes of blazars. In FSRQs, radio and soft X-ray emissions are tightly correlated, while in BL Lacs the relationship is much more scattered.

The results we obtained ({Table }\ref{tab:xray_radio}) confirm that FSRQs are in an accretion-driven regime, which fuels both the radio jet and the disc emission. On the other hand, in BL Lacs, the jets are mainly rotation driven. 
However, alternative factors could contribute to the observed trends. For instance, BL Lacs exhibit a higher variability in the X-ray band, which may introduce additional scatter in the correlation \citep{Ghisellini2017}. 

Moreover, the impact of the circumnuclear medium could play a role in shaping the X-ray emission properties, particularly in FSRQs, where disc-related processes contribute significantly to the total luminosity.
A deeper understanding of these effects will require time domain studies incorporating simultaneous X-ray and radio observations.

When comparing the X-ray luminosity with the 20 GHz emission, it is essential to account for selection bias. The sources in our sample that have both a 20 GHz and an X-ray counterpart represent a highly efficient subset of our data. Consequently, the tighter correlation between X-ray and radio emissions is not surprising given that the fraction of FSRQs is higher at these frequencies compared to lower frequencies. Deeper millimetre observations are required to gain a comprehensive scientific understanding of the behaviour at higher frequencies.
\section{Summary and future perspectives}\subsection{Summary}
In this study, we have explored the radio properties of 4FGL blazars detected in the FLASH continuum survey while focusing on their spectral characteristics and multi-wavelength correlations. The main findings are summarised as follows:\begin{itemize}
    \item {Radio SED characteristics:} About 50\% of the sources in the sample exhibit a clearly re-triggered peaked spectrum, while for approximately 14\% of them, such a spectrum can be safely excluded. This suggests that the emission is dominated by a single emitting region rather than by the superposition of multiple compact, self-absorbed regions.
    \item {Spectral trends:} The median peak frequency ($\nu_{\text{peak}}$) is approximately 4.6 GHz, with 55\% of sources showing peaks above 10 GHz (high-frequency peakers), indicating a population of likely young or recently re-activated blazars.
    \item {Extended emission:} At low frequencies (e.g. below 100 MHz), the emission is characterised by a spectral index of $\sim-$0.54, which is consistent with the presence of an additional steep-spectrum emission that was previously missed due to a lack of high-resolution and deep low-frequency data.
    \item {Radio-gamma correlation:} A strong correlation between radio and gamma-ray luminosities was identified, particularly at 855 MHz, emphasising the dominant contribution of relativistic jets to the observed multi-wavelength emission.
    \item {Radio--X-ray correlation:} The X-ray luminosity (0.2--2.0 keV) is tightly correlated with radio luminosity for FSRQs, reflecting their accretion-driven nature, while BL Lacs show a more scattered relationship, which is consistent with their jet-driven dynamics.
    \item {Blazar subclass differences:} FSRQs display more efficient high-energy pumping and dominate the bright end of the radio and gamma-ray luminosity spectrum, while BL Lacs exhibit greater variability in spectral behaviour, reflecting their evolutionary differences.
\end{itemize}
\subsection{Future perspectives}
Building on these findings, future studies will be aimed at deepening our understanding of the physical and environmental properties of blazars. 
As FLASH is an ongoing survey, we will expand the blazar sample to include the full five-year dataset upon its completion.
Further, a key next step will be to analyse the spectroscopic properties of these objects using FLASH, XMM-Newton, and optical lines data (e.g. broad-line region emission). This will help us provide detailed insights into the physics of blazars and their environments, and offer clues about the re-triggering mechanism and the dual nature of their radio emission. Finally, leveraging instruments such as the SKA Observatory and Cherenkov Telescope Array will allow for broader frequency coverage and improved sensitivity and thus enable the detection of faint sources and the study of synchrotron emission at unprecedented depth.
\section{Data availability}
Tables listing the properties of the sources are only available in electronic form 
at the CDS via anonymous ftp to \texttt{cdsarc.u-strasbg.fr} (130.79.128.5) or via
\url{http://cdsweb.u-strasbg.fr/cgi-bin/qcat?J/A+A/aa54523-25}.
.

    
    
    
    
\begin{acknowledgements}
This work was carried out in part during an Erasmus+ exchange program at CSIRO ATNF in Marsfield, Sydney. MB gratefully acknowledges the support provided by Erasmus+ and the CSIRO ATNF for facilitating this research opportunity. This scientific work uses data obtained from Inyarrimanha Ilgari Bundara, the CSIRO Murchison Radio-astronomy Observatory. We acknowledge the Wajarri Yamaji People as the Traditional Owners and native title holders of the Observatory site. CSIRO’s ASKAP radio telescope is part of the Australia Telescope National Facility (https://ror.org/05qajvd42). Operation of ASKAP is funded by the Australian Government with support from the National Collaborative Research Infrastructure Strategy. ASKAP uses the resources of the Pawsey Supercomputing Research Centre. Establishment of ASKAP, Inyarrimanha Ilgari Bundara, the CSIRO Murchison Radio-astronomy Observatory and the Pawsey Supercomputing Research Centre are initiatives of the Australian Government, with support from the Government of Western Australia and the Science and Industry Endowment Fund. 
This work was partially funded from the projects: INAF GO-GTO Normal 2023 funding scheme with the project 'Serendipitous H-ATLAS-fields Observations of Radio Extragalactic Sources (SHORES)'. MVZ acknowledges partial financial support from the 'Fondazione CR Firenze' with the program 'Ricercatori a Firenze 2023'. The colour palettes used in this paper were directly extracted from Claude Monet's paintings \textit{Water Lilies} and \textit{San Giorgio Maggiore at Dusk}.
\end{acknowledgements}
\bibliographystyle{aa}
\bibliography{PASPsample631}
\begin{appendix}
\onecolumn

\includepdf[
  nup=3x4,
  pages=1-12,
  pagecommand={
    \section{Spectral Energy Distributions}\label{sec:appendix}
 Fig. A.1. Spectral Energy Distributions of the blazars in the sample.
    \vspace{0.5cm}
  }
]{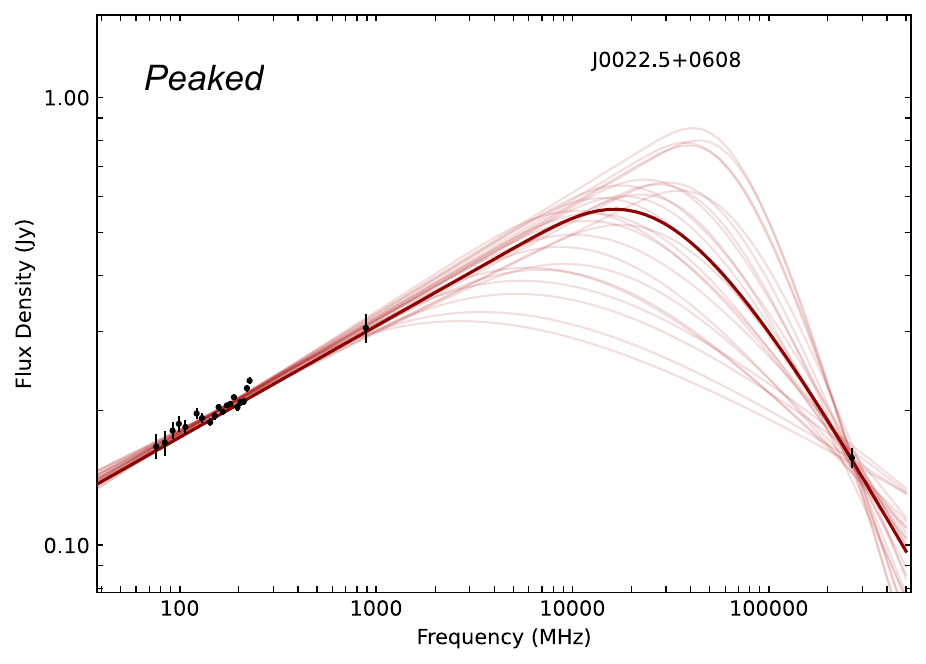}

\includepdf[
  nup=3x4,
  pages=12-,
  pagecommand={\thispagestyle{plain}\centering Fig. A.1. Continued.}
]{all_seds.pdf}


\end{appendix}



\end{document}